\documentclass[preprint,showpacs,amsmath,aps]{revtex4}
\usepackage{multirow}
\usepackage{amsmath}
\usepackage{epsfig}
\usepackage{subfigure}
\usepackage{float}
\usepackage{verbatim} 
\usepackage{booktabs}

\def\met{\ifmmode%
\setbox0=\hbox{$E_t$}%
\setbox1=\hbox to\wd0{\hss$/$\hss}\else%
\setbox0=\hbox{E_t}%
\setbox1=\hbox to\wd0{\hss/\hss}\fi%
E_t\hskip-\wd0\box1 }

\def\tr{\mathop{\rm tr}\nolimits}
\def\bentarrow{\:\raisebox{1.3ex}{\rlap{$\vert$}}\!\rightarrow}
\def\bothdk#1#2#3#4#5{
\begin{array}{r c l}
#1 & \to & #2#3 \\
 & & \:\raisebox{1.3ex}{\rlap{$\vert$}}\raisebox{-0.5ex}{$\vert$}%
\phantom{#2}\!\bentarrow #4 \\
 & & \bentarrow #5
\end{array}
}

\begin{document}

\bigskip
\title{Hadronic production of a Higgs boson and two jets
at next-to-leading order}

\author{John M. Campbell and R. Keith Ellis}
\affiliation{
Theory Department, Fermi National Accelerator Laboratory,\\
P.O. Box 500, Batavia, IL 60510, USA}

\author{Ciaran Williams}
\affiliation{Department of Physics, University of Durham, \\
Durham, DH1 3LE, United Kingdom}

\preprint{
FERMILAB-PUB-10-013-T,
IPPP/10/06}

\begin{abstract}
We perform an update of the next-to-leading order calculation of the rate for Higgs boson production in association with two jets.
Our new calculation incorporates the full analytic result for the one-loop virtual amplitude. This new theoretical information 
allows us to construct a code including the decay of the Higgs boson without incurring a prohibitive penalty in computer running time.
Results are presented for the Tevatron, where implications for the Higgs search are sketched, and also for a range
of scenarios at the LHC.
\end{abstract}


\maketitle

\section{Introduction}

In the coming years the hadron colliders at
Fermilab and CERN will focus on the hunt for the Higgs boson. The large data sample currently being collected at the Tevatron
will certainly lead to improved limits on the Higgs mass~\cite{Collaboration:2009je}, or even
evidence for its existence. The Large Hadron
Collider has the potential to confirm the existence of the Higgs
boson~\cite{Ball:2007zza,Aad:2009wy} between the lower limit set by LEP~\cite{Collaboration:2009jr} and the
upper bound suggested by perturbative unitarity~\cite{Lee:1977eg,Lee:1977yc}.

Such claims are based on detailed analyses that clearly require reliable theoretical
predictions for the production cross sections and characteristics. It is well-known that
leading order predictions for such quantities, based on tree-level Feynman diagrams alone,
are not sufficiently trustworthy for this purpose. The calculations are plagued by large
uncertainties in their overall normalization and moreover, important kinematic effects
are often missed.

In this paper we present results for the production of a Higgs boson in association with two
jets. Our calculation is performed at next-to-leading order (NLO) using an effective Lagrangian to express the coupling of gluons to 
the Higgs field~\cite{Wilczek:1977zn},
\begin{equation} \label{EffLag}
\mathcal{L}_H^{\mathrm{int}} = \frac{C}{2} \, H\,\tr
G_{\mu\nu}\,G^{\mu\nu}\, . 
\end{equation}
where the trace is over the color degrees of freedom. 
At the order required in this paper, the coefficient $C$  
is given in the $\overline{\rm MS}$ scheme by~\cite{Djouadi:1991tka,Dawson:1990zj},
\begin{equation}
C =\frac{\alpha_S}{6 \pi v} \Big( 1 +\frac{11}{4 \pi} \alpha_S\Big)
 + {\cal O}(\alpha_S^3) \;.
\end{equation}
Here $v$ is the vacuum expectation value of the Higgs field, $v = 246$ GeV. 

This Lagrangian replaces the full one-loop coupling of the Higgs boson to the gluons
via an intermediate top quark loop by an effective local operator.
The effective Lagrangian approximation is valid in the limit $m_H < 2 m_t$ and, in the presence of additional jets,
when the transverse momenta of the jets  is not much larger than the top mass $m_t$~\cite{DelDuca:2001fn}.
A commonly used improvement of the effective Lagrangian 
approximation is to multiply the resulting differential jet cross section by a ratio $R$ given by,
\begin{equation}
R = \frac{\sigma_{\rm finite~m_t}(gg \to H)}{\sigma_{m_t \to \infty}(gg \to H)} \;,
\label{Reqn}
\end{equation}
where $\sigma (gg \to H)$ is the total cross section.
Setting  $x = 4 m_t^2/m_H^2$ the correction for the finite mass of the top quark in the region
$x > 1$ is~\cite{Djouadi:1991tka},
\begin{equation} \label{finitemassR}
R=\Bigg[\frac{3 x}{2} \Big( 1-(x-1) \Big[\sin^{-1}\frac{1}{\sqrt{x}}\Big]^2\Big)\Bigg]^2 \;.
\end{equation}
This rescaling is known to be an excellent approximation for the Higgs + 2 jet rate, see Ref.~\cite{DelDuca:2001fn} and references
therein. However for the case of Higgs + 1 jet it has been found that the effect of bottom quark loops and additional electroweak
diagrams can also be important~\cite{Keung:2009bs} and these effects should also be included. Our numerical results for the 
Higgs cross section will not include the rescaling of Eqs.~(\ref{Reqn},\ref{finitemassR}).

\section{New features of this paper}

The phenomenology of the production of a Higgs boson in association with two jets 
has been presented in Ref.~\cite{Campbell:2006xx} for the case of the LHC operating at $\sqrt s = 14$~TeV.
The NLO analysis in that paper was based
on real matrix elements for the Higgs+5 parton processes given in Ref.~\cite{DelDuca:2004wt}, supplemented  
by the results of Ref.~\cite{Dixon:2004za,Badger:2004ty} in the cases where these latter results 
lead to more efficient code.  
In Ref.~\cite{Campbell:2006xx} the virtual matrix element corrections for the Higgs + 4 parton process were taken from 
Ref.~\cite{Ellis:2005qe}. For the $Hgggg$ and $Hq\bar{q}gg$ sub-processes the virtual corrections were based 
on a semi-numerical technique~\cite{Ellis:2005zh}, whilst the matrix elements squared 
for the one-loop processes $Hq\bar{q} q^\prime\bar{q}^\prime $ and $Hq\bar{q} q\bar{q}$ 
were given analytically in Ref.~\cite{Ellis:2005qe}.

In the three years since Ref.~\cite{Campbell:2006xx} was published a great deal of effort 
has been devoted to the {\it analytic} calculation of one-loop corrections to Higgs + $n$-parton
amplitudes, with particular emphasis on the $n=4$ amplitudes which are relevant for this study.
The complete set of one-loop amplitudes for all Higgs + 4 parton processes
is now available and analytic expressions can be found in the following references:
\begin{itemize}
\item $Hgggg$: Refs.~\cite{Berger:2006sh,Badger:2006us,Badger:2007si,Glover:2008ffa,Badger:2009hw};
\item $H\bar q q g g$: Refs.~\cite{Dixon:2009uk,Badger:2009vh};
\item $H\bar q q \bar Q Q$: Ref.~\cite{Dixon:2009uk}.
\end{itemize}
These new analytic results have now been included in the MCFM package, version 5.7
(which  may be downloaded from {\tt mcfm.fnal.gov}), 
leading to a considerable improvement in the speed of the code.
For the processes involving two quark-antiquark pairs,
the matrix elements squared given in Ref.~\cite{Ellis:2005qe} are implemented in MCFM,
rather than the amplitudes of Ref.~\cite{Dixon:2009uk}, because they lead to faster code.
The values of the amplitudes calculated by the new analytic code and the previous semi-numerical
code~\cite{Campbell:2006xx} are in full numerical agreement for all amplitudes.

The improvement in the performance of our numerical code means that it is appropriate to revisit the phenomenology 
of Higgs + 2 jet production and to extend it in a number of ways.  The improvement in the speed of the code means 
that it is possible to include the decays of the Higgs boson, specifically for the processes:
\begin{equation}
h_1+h_2 \to H+j_1+j_2 \to \tau^+ + \tau^- + j_1+j_2
\label{htt}
\end{equation}
\begin{equation}
h_1+h_2 \to H+j_1+j_2 \to b + \bar b + j_1+j_2
\label{hbb}
\end{equation}
\begin{equation}
\bothdk{h_1+h_2 \to H+j_1+j_2}{W^- +}{W^+ + j_1+j_2}{\nu+e^+}{e^-+{\bar \nu}}
\label{hww}
\end{equation}
\begin{equation}
\bothdk{h_1+h_2 \to H+j_1+j_2}{Z + }{Z + j_1+j_2}{e^- + e^+}{\mu^- + \mu^+}
\label{hzz}
\end{equation}
where $h_1$, $h_2$ represent partons inside the incident hadron beams. All four of these processes are included
in MCFM v5.7.

\section{Parameters}
Throughout this paper we make use of the MSTW2008 parton distribution functions~\cite{Martin:2009iq},
using the LO fit ($\alpha_s(M_Z) = 0.13939$ and 1-loop running) for the lowest order calculation and the
NLO fit ($\alpha_s(M_Z) = 0.12018$ and 2-loop running) at NLO.
The $W$ mass and width are chosen to be,
\begin{equation}
m_W = 80.398~{\rm GeV}, \qquad \Gamma_W = 2.1054~{\rm GeV} \;.
\label{wparams}
\end{equation}
The mass is taken from Ref.~\cite{Amsler:2008zzb}. The total width given in Eq.~(\ref{wparams}) is derived from the
measured branching
ratio for $W \to \ell \bar\nu$, $10.80 \pm 0.09\%$~\cite{Amsler:2008zzb} by using a lowest order calculation
of the partial width,
\begin{equation}
\Gamma( W \to \ell \bar \nu) = \frac{G_F}{\sqrt 2} \frac{m_W^2}{6 \pi} \;.
\end{equation}
This ensures that our calculation incorporates the best possible value for the $W$ branching ratio which is
determined to about 1\%.
The values of the total Higgs width are taken from the program {\tt hdecay}~\cite{Djouadi:1997yw}, version 
3.51.

To define the jets we perform clustering according to the $k_T$ algorithm~\cite{Blazey:2000qt}, with
jet definitions detailed further below.

\section{Tevatron results}
We use a very simple set of inclusive cuts, with no requirements on the Higgs boson decay
products,
\begin{equation} \label{tevjetcuts}
p_t({\rm jet})>15\;{\rm GeV}, \qquad
|\eta_{\rm jet}|<2.5, \qquad
R_{{\rm jet},{\rm jet}}>0.4 \; .
\end{equation}

\begin{table}
\begin{tabular}{|c||c|c||c|c|}
\hline
& \multicolumn{2}{|c||}{$|\eta_{\rm jet}| < 2.5$} & \multicolumn{2}{|c|}{$|\eta_{\rm jet}| < 2$}  \\
\hline
Process               & $\sigma_{LO}$~[fb]  & $\sigma_{NLO}$~[fb] & $\sigma_{LO}$~[fb]  & $\sigma_{NLO}$~[fb] \\
\hline
Higgs + 0 jets        & $1.25$              & $1.98$ &  $1.25$  	    & $2.05$\\
Higgs + 1 jets        & $0.84$              & $1.16$ &  $0.74$  	    & $1.07$\\
Higgs + $\geq$ 2 jets & $0.35$              & $0.48$ &  $0.28$  	    & $0.39$\\
\hline
\end{tabular}
\caption{Cross section for Higgs + jet production and decay into $W^-(\to \mu^- \bar{\nu}) W^+(\to \nu e^+)$
at $\sqrt{s}=1.96$~TeV for $M_H=\mu=160$~GeV. In the second and third columns, only the cuts of Eq.~(\ref{tevjetcuts})
are applied. For the results in the final two columns the more stringent cut, $|\eta_{\rm jet}|<2$ is applied,
in order to allow a comparison with Ref.~\cite{Anastasiou:2009bt}.
\label{tevjetresults} }
\end{table}
At the Tevatron the search for the Higgs boson has been divided into jet bins. To set the stage for this we show in 
Table~\ref{tevjetresults} the expected cross section in each bin due to the gluon fusion mechanism.
The parameter $\mu$ is the renormalization and factorization scale, which we set equal to $m_H$ here.
We note that next-to-next-to-leading order (NNLO) results for the Higgs + 0 jet cross section are given in~\cite{Anastasiou:2009bt},
based on the earlier calculations in Refs.~\cite{Anastasiou:2005qj,Anastasiou:2007mz,Grazzini:2008tf}.
From table~\ref{tevjetresults} columns 3 and 5, we see that the Higgs~+~$\geq$~2~jets bin constitutes about
13\% of the cross section for $|\eta_{\rm jet}|<2.5$ and 11\% with $|\eta_{\rm jet}|<2$.

It is interesting to compare the number for the fraction of Higgs~$+ \ge 2$ jet events ($|\eta_{\rm jet}|<2$) with the percentage
extracted from Table 2 of~\cite{Anastasiou:2009bt}, which is quoted as $4.9$\%.
Our number is deficient in that it does not include NNLO corrections
to the Higgs + 0 jet rate. Our calculation treats all jet bins consistently at NLO. The inclusion of the NNLO correction to the
Higgs + 0 jet bin will reduce our number. On the other
hand, the calculation of Ref.~\cite{Anastasiou:2009bt} is deficient because it does not treat all bins consistently at NNLO, i.e.
it does not include NNLO corrections for the Higgs~$+ 1$ jet rate or NLO+NNLO effects for the Higgs~$+ \ge 2$ jet rate.
We roughly estimate that including the NNLO effects in the Higgs~$+ 0$ jet bin would move our central
value from $11$\% to $10$\%. Overall, because the corrections are quite substantial, the theoretical estimate of the fraction of
events in the Higgs~$+ \ge 2$~jet bin is quite uncertain. 

Despite the fact that the fraction of events in the Higgs $+ \ge 2$ jet bin is small, it is important because
the associated uncertainty is large. 
We investigate this issue in Table~\ref{tevresults}, where we give the cross section for the process of Eq.~(\ref{hww}) using a
selection of values for the Higgs mass of current interest for the Tevatron.
\begin{table}
\begin{tabular}{|c|c|c|c|c|c|}
\hline
$m_H$~[GeV]      & 150     &  160     & 165    & 170    & 180 \\
$\Gamma_H$~[GeV] & 0.0174  &  0.0826  & 0.243  & 0.376  & 0.629 \\
\hline
$\sigma_{LO}$~[fb]  &$0.329^{+92\%}_{-45\%}$&$0.345^{+92\%}_{-44\%}$&$0.331^{+92\%}_{-44\%}$&$0.305^{+92\%}_{-44\%}$&$0.245^{+91\%}_{-44\%}$ \\
$\sigma_{NLO}$~[fb] &$0.447^{+37\%}_{-30\%}$&$0.476^{+35\%}_{-31\%}$&$0.458^{+36\%}_{-31\%}$&$0.422^{+41\%}_{-30\%}$&$0.345^{+37\%}_{-31\%}$ \\
\hline
Finite $m_t$ correction, $R$ & $1.098 \pm 0.003 $&$ 1.113 \pm 0.003 $&$ 1.122 \pm 0.004 $&$ 1.130 \pm 0.005 $&$ 1.149 \pm0.005 $ \\ 
\hline
\end{tabular}
\caption{Cross section for Higgs + 2 jet production and decay into $W^-(\to \mu^- \bar{\nu}) W^+(\to \nu e^+)$
at $\sqrt{s}=1.96$~TeV. Only the cuts of Eq.~(\ref{tevjetcuts}) are applied. The correction factor for each Higgs mass,
given by Eq.~(\ref{finitemassR}), is also shown.
\label{tevresults}}
\end{table}
In the table we give the results for the leading order and next-to-leading order
cross sections, calculated using LO and NLO MSTW2008 PDFs respectively. For the range of Higgs masses considered, the QCD corrections
increase the cross section by approximately 40\% (for the central value, $\mu = m_H$). The theoretical error is estimated
by varying the common renormalization and factorization scale in the range, $m_H/2 < \mu < 2m_H$. As can be seen from
the table, even though including the next-to-leading order corrections leads to a considerable improvement in the theoretical
error, the remaining  error is still quite sizeable. We do not include a factor to correct for the finite top mass, but in order
to facilitate comparison with other calculations we also tabulate this factor $R$ (computed using Eq.~(\ref{finitemassR})) using
a value for the top quark mass of $m_t=172.5 \pm 2.5$~GeV.

In the spirit of Ref.~\cite{Anastasiou:2009bt}, we can now estimate the theoretical uncertainty on the number of
Higgs signal events originating from gluon fusion. By using the fractions of the Higgs cross section in the
different multiplicity bins taken from Ref.~\cite{CDFnote9500}, we can update Eq.~(4.3) of Ref.~\cite{Anastasiou:2009bt}
(for a Higgs boson of mass $160$~GeV) with,
\begin{equation}
\frac{\Delta N_{\rm signal}({\rm scale})}{N_{\rm signal}} =  
 60\% \cdot \left({^{ +5\%}_{ -9\%}} \right) 
+29\% \cdot \left({^{+24\%}_{-23\%}} \right) 
+11\% \cdot \left({^{+35\%}_{-31\%}} \right) = \left({^{+13.8\%}_{-15.5\%}} \right)  
\end{equation}
Only the uncertainty on the Higgs~$+ \ge 2$ jet bin has been modified, using the results from Table~\ref{tevresults}.
The corresponding determination using the LO uncertainty in the Higgs~$+\ge 2$ jet bin is
$(+20,0 \%, -16.9 \%)$~\cite{Anastasiou:2009bt}, so this represents a modest improvement in the overall theoretical error.   

The correspondence of our results with those of Anastasiou et al. is somewhat obscured by the fact that the total Higgs width used
in Ref.~\cite{Anastasiou:2009bt} is about $7$\% smaller at $m_H =160$~GeV than the value
given in our Table~\ref{tevresults}. Taking this fact into account and including the
finite top mass correction tabulated in Table~\ref{tevresults} we find that our
NLO Higgs + 1~jet and LO Higgs + 2~jet cross sections in Table~\ref{tevjetresults}
are in agreement with the corresponding numbers ($1.280$ and $0.336$~fb) from Table~2 of Ref.~\cite{Anastasiou:2009bt}.

\subsection{Effect of additional search cuts}
We also investigate the behaviour of the LO and NLO predictions in the kinematic region
relevant for the latest Tevatron Higgs exclusion limits. Therefore,  
in addition to the jet cuts above, we also consider cuts on the decay products of the
$W/W^\star$ that are produced by the Higgs boson. These cuts correspond very closely
to a recent CDF analysis~\cite{CDFnote9887}, although the treatment of lepton acceptance is
simplified.

\begin{itemize}
\item One of the leptons from the $W$ decays (the ``trigger'' lepton, $\ell_1$) is required to be relatively
hard and central, $p_t^{\ell_1}>20\;{\rm GeV}$, $|\eta^{\ell_1}|<0.8$ whilst the other ($\ell_2$)
may be either softer or produced at slightly higher pseudorapidity,
$p_t^{\ell_2}>10\;{\rm GeV}$, $|\eta^{\ell_2}|<1.1$.
\item The invariant mass of the lepton pair is bounded from below (to eliminate virtual photon contributions),
$m_{\ell_1 \ell_2} > 16\;{\rm GeV}$.
\item Each lepton must be isolated. Any jet found by the algorithm that lies within a $\eta-\phi$ distance of $0.4$
from a lepton should have a transverse momentum less than $10\%$ of that of the lepton itself.
\item The missing transverse momentum -- in our parton level study, the sum of the two neutrino momenta -- is
constrained using the $\met^{\rm spec}$ variable defined by~\cite{CDFnote9887},
\begin{equation}
\met^{\rm spec} = \met \sin \left[ {\rm min} \left( \Delta \phi , \, \frac{\pi}{2} \right) \right] \;.
\end{equation}
$\Delta \phi$ is the distance between the $\met$ vector and the nearest lepton or jet.
We require that $\met^{\rm spec} > 25 \;{\rm GeV}$.
\end{itemize}

\begin{figure}
\includegraphics[width=12.5cm]{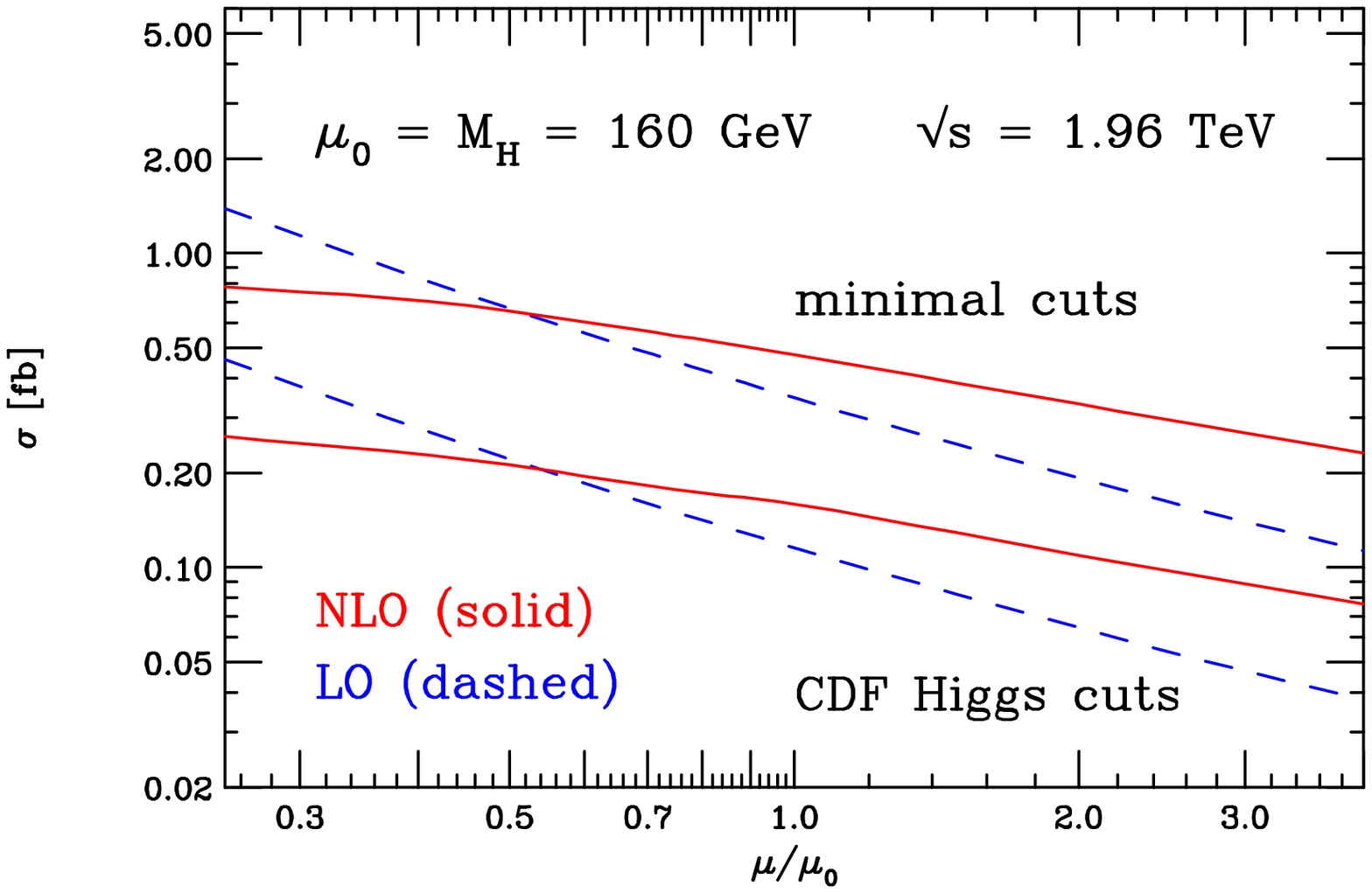}
\caption{Scale dependence for the Higgs + 2 jet cross section, with the Higgs decay into $W^-(\to \mu^- \bar{\nu}) W^+(\to \nu e^+)$,
at the Tevatron and using the a central scale $\mu_0 = M_H$.
Results are shown for the minimal set of cuts in Eq.~(\ref{tevjetcuts}) (upper curves) and for cuts that mimic the latest
CDF $H \to WW^\star$ analysis (lower curves).\label{mudeptev}} 
\end{figure}

In Figure~\ref{mudeptev} we see the scale dependence of the LO and NLO cross sections for $m_H = 160$~GeV.
The upper two curves show the case of the minimal set of cuts in Eq.~(\ref{tevjetcuts}) and the lower curves show
the results when including the Higgs search cuts above.
Applying the additional cuts on the Higgs decay products does not change the scale dependence, indicating
that the isolation and missing transverse momentum cuts (that are sensitive to additional radiation) do not play an important role.
Applying the additional search cuts does not alter the behaviour of the NLO prediction in the Higgs~$+ \ge 2$ jet bin, so that the 
results presented in the previous section (with no cuts on the Higgs decay products) are sufficient to estimate the
percentage theoretical uncertainty.

\section{LHC results}

In order to study the impact of the NLO corrections at the LHC, we adopt a different set of cuts to define the jets.
The rapidity range of the detectors is expected to be much broader, allowing for a larger jet separation too, and we choose a
somewhat higher minimum transverse momentum,
\begin{equation}
p_t({\rm jet})>40\;{\rm GeV}, \qquad
|\eta_{\rm jet}|<4.5, \qquad
R_{{\rm jet},{\rm jet}}>0.8 \; .
\label{lhcjetcuts}
\end{equation}
In this section we do not consider the decay of the Higgs boson for the sake of simplicity.

\begin{figure}
\includegraphics[width=12.5cm]{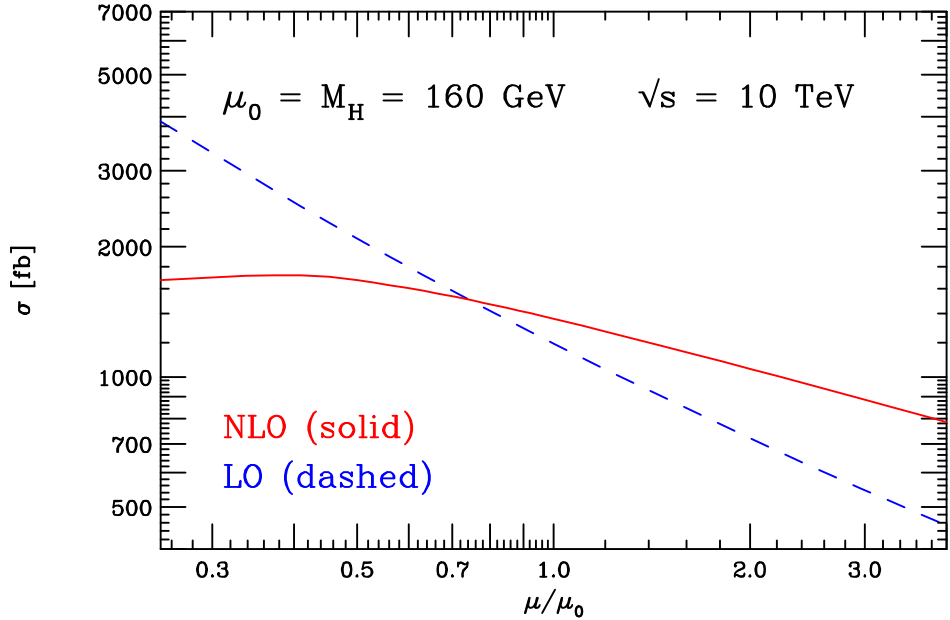}
\caption{Scale dependence for the Higgs boson + 2 jet cross section, using the basic set of cuts in Eq.~(\ref{lhcjetcuts}) and
a central scale choice $\mu_0 = m_H$. \label{mudeplhc}}
\end{figure}

Since results for this scenario have already been discussed at some length~\cite{Campbell:2006xx}, we restrict ourselves
to a short survey of the essential elements of the phenomenology at the lower centre-of-mass energy, $\sqrt s = 10$~TeV.
We present the scale dependence of the LHC cross section for Higgs + 2 jets ($m_H=160$~GeV) in Figure~\ref{mudeplhc}.
We have also checked the agreement of our calculation with previous results~\cite{Campbell:2006xx} at $\sqrt s = 14$~TeV,
taking into account the different choice of parton distribution functions used in that reference.
As noted in the earlier paper~\cite{Campbell:2006xx}, the corrections are quite modest using our central scale choice, $\mu_0 = \mu_H$,
increasing the cross section by approximately 15\%. Once again, although the scale dependence is much reduced it is still substantial.

For the sake of illustration we have chosen $m_H=160$~GeV in the study above. To illustrate the effect of the QCD corrections
more broadly, in Table~\ref{lhcmhresults} we give the cross sections for Higgs masses in the range $120$~GeV~$ < m_H < 200$~GeV.
\begin{table}
\begin{tabular}{|c|c|c|c|c|c|}
\hline
$m_H$~[GeV]      & 120     &  140     & 160    & 180    & 200 \\
$\Gamma_H$~[GeV] & 0.0036  &  0.0083  & 0.0826  & 0.629  & 1.426 \\
\hline
$\sigma_{LO}$~[pb]  &$ 1.88^{+78\%}_{-40\%}$&$ 1.48^{+76\%}_{-40\%}$&$ 1.20^{+75\%}_{-40\%}$&$ 0.98^{+74\%}_{-39\%}$&$ 0.81^{+73\%}_{-39\%}$ \\
$\sigma_{NLO}$~[pb] &$ 1.98^{+20\%}_{-23\%}$&$ 1.63^{+22\%}_{-23\%}$&$ 1.36^{+23\%}_{-23\%}$&$ 1.15^{+24\%}_{-23\%}$&$ 0.98^{+25\%}_{-24\%}$ \\
\hline
Finite $m_t$ correction, $R$ & $1.060 \pm 0.002 $&$ 1.084 \pm 0.003 $&$ 1.113 \pm 0.004 $&$ 1.149 \pm 0.005 $&$ 1.191 \pm0.007 $ \\ 
\hline
\end{tabular}
\caption{Cross section and uncertainties for Higgs + 2 jet production at $\sqrt{s}=10$~TeV with the cuts of Eq.~(\ref{lhcjetcuts}).
 The correction factor for each Higgs mass, given by Eq.~(\ref{finitemassR}), is also shown.
\label{lhcmhresults}}
\end{table}
It is within this range that the Higgs + 2 jet process considered here is of most interest, due to its interplay with the electroweak
weak boson fusion channel. We observe that the effect of the QCD corrections increases from about $5\%$ for $m_H=120$~GeV to
$21\%$ for $m_H=200$~GeV.
Estimating the theoretical error in the same way as before, we see that the uncertainty is slightly less at the LHC than at the Tevatron.

It is also interesting to consider the dependence of the cross section on the minimum transverse momentum required
for the observed jets. Results for several other values of this threshold, either side of our default value of
$40$~GeV, are shown in Table~\ref{lhcptresults}. As can be seen from the table, the percentage effect of the NLO corrections
on the total rate is practically independent of the value of $p_t^{\rm min}({\rm jet})$ in the range studied.
\begin{table}
\begin{tabular}{|c|c|c|c|c|c|}
\hline
$p_t^{\rm min}({\rm jet})$~[GeV] & 20 & 25 & 30 & 40 & 50 \\
\hline
$\sigma_{LO}$~[pb]   & 3.66 & 2.62 & 1.96 & 1.20 & 0.79  \\
$\sigma_{NLO}$~[pb]  & 4.17 & 3.02 & 2.26 & 1.36 & 0.88  \\
\hline
\end{tabular}
\caption{Cross section for Higgs + 2 jet production at $\sqrt{s}=10$~TeV, with $m_H=160$~GeV and the minimum
jet $p_t$ allowed to vary from that specified in Eq.~(\ref{lhcjetcuts}).
\label{lhcptresults}} 
\end{table}

\subsection{Weak boson fusion}
As noted above, the process studied in this paper produces the same final state as expected from Higgs production
via weak boson fusion (WBF). Although the electroweak process is expected to dominate once appropriate search cuts are
employed, the remaining fraction of events originating from gluon fusion must be taken into account when considering
potential measurements of the Higgs coupling to $W$ and $Z$ bosons.

\begin{figure}
\includegraphics[width=12.5cm]{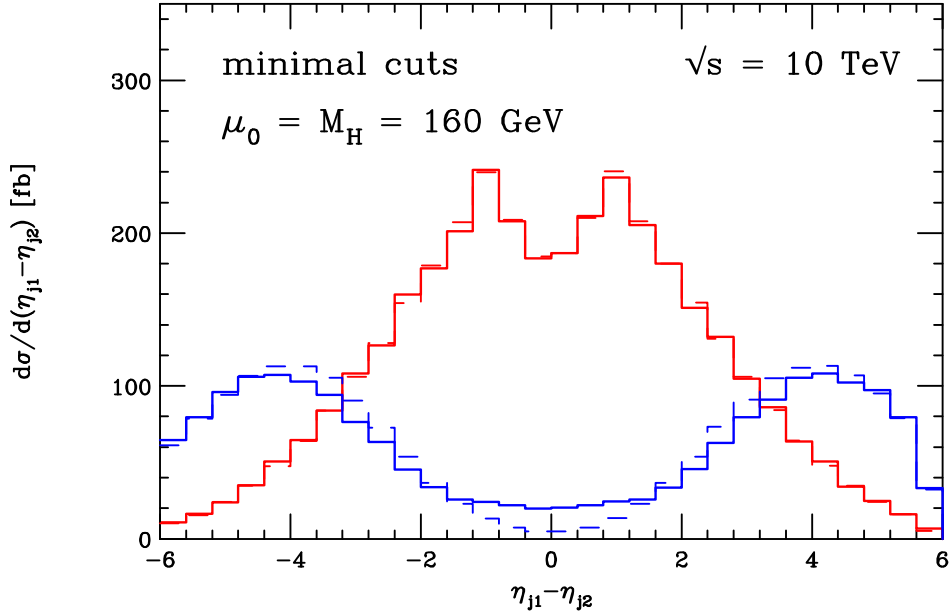}
\caption{The jet pseudorapidity difference in gluon fusion (red) and weak boson fusion (blue). The NLO predictions are shown
as solid histograms, while the dashed lines indicate the LO predictions normalized to the corresponding NLO cross sections. 
\label{jetdiff}}
\end{figure}

To address this issue, in this section we present a brief study of the rate of events expected using typical weak
boson fusion search cuts. In addition to the cuts already imposed (Eq.~(\ref{lhcjetcuts})), these correspond to,
\begin{equation}
\left| \eta_{j_1} - \eta_{j_2} \right| > 4.2 \;, \qquad \eta_{j_1} \cdot \eta_{j_2} < 0 \;,
\label{wbfcuts}
\end{equation}
where $j_1$ and $j_2$ are the two jets with the highest transverse momenta. These cuts pick out the distinctive signature
of two hard jets in opposite hemispheres separated by a large distance in pseudorapidity. This is illustrated in
Fig,~\ref{jetdiff}, where we compare the distributions of the jet pseudorapidity difference (without these cuts) in
both gluon fusion and weak boson fusion. We note in passing that the shape of this distribution for the
weak boson fusion process is slightly altered at NLO, whilst the shape of the prediction for the gluon fusion process is
essentially unchanged.

\begin{figure}
\includegraphics[width=12.5cm]{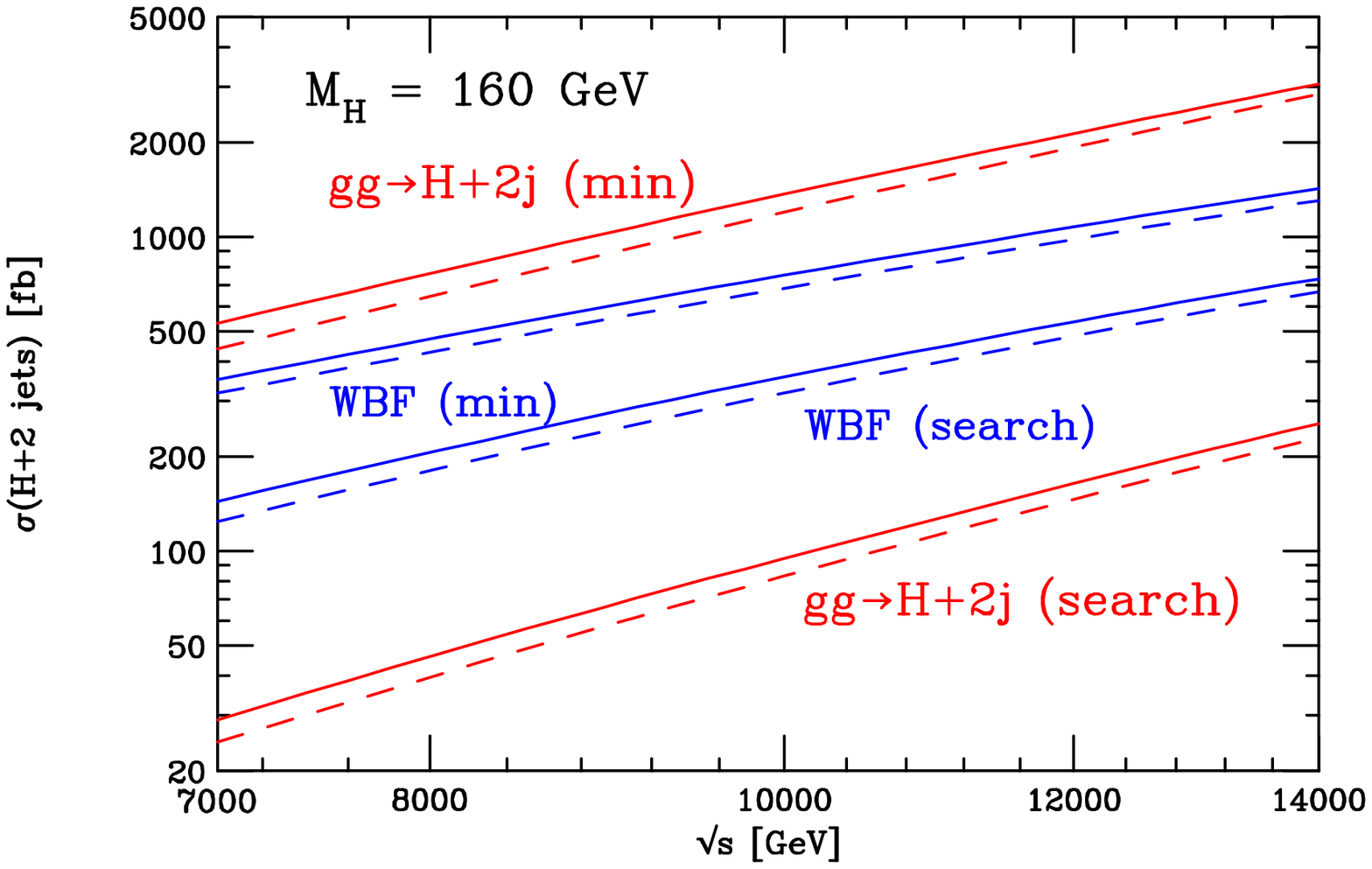}
\caption{The $\sqrt s$ dependence of the cross section for $m_H=160$~GeV at LO (dashed) and NLO (solid). Results are shown for
the minimal set of cuts in Eq.~(\ref{lhcjetcuts}) (two upper red curves) and after application of the additional WBF Higgs search cuts
given in Eq.~(\ref{wbfcuts}) (two lower red curves).
The cross section for the weak boson fusion process is also shown for comparison (four central blue curves).
\label{rtsdep}}
\end{figure}

In figure~\ref{rtsdep} we show the dependence of the cross section on the c.o.m. energy, from $\sqrt s = 7$~TeV (corresponding to
the initial running in 2010-11) to $\sqrt s =14$~TeV (design expectations).
We show the cross section both before and after application of the additional weak boson fusion search cuts given
in Eq.~(\ref{wbfcuts}), together with the corresponding results for the WBF process (also calculated using
MCFM~\cite{Berger:2004pc}). The QCD corrections to both processes decrease slightly as $\sqrt s$ is increased, whilst the
ratio of the gluon fusion to WBF cross sections after the search cuts are applied increases from $20$\% at $7$~TeV
to $35\%$ at $14$~TeV. This indicates that, viewed as a background to the weak boson fusion process, the hadronic
Higgs + 2 jet process is less troublesome at energies below the nominal design value.

\section{Conclusions}

In this paper we have presented phenomenological predictions for the production of a Higgs boson and two jets
through gluon fusion. These predictions have been made possible through the implementation of recent compact analytic
results for the relevant 1-loop
amplitudes~\cite{Berger:2006sh,Badger:2006us,Badger:2007si,Glover:2008ffa,Badger:2009hw,Dixon:2009uk,Badger:2009vh}.
The speed with which these amplitudes can be evaluated has enabled us to improve upon an existing
semi-numerical implementation of the same process~\cite{Campbell:2006xx}, with various decays of the Higgs boson
now included.

We have investigated the behaviour of the NLO cross section at the Tevatron, where contributions from this channel
form part of the event sample for the latest Higgs searches~\cite{Collaboration:2009je}. We find that corrections
to the event rate in the Higgs~$+ \ge 2$~jet bin are modest and that the estimate of the theoretical error is reduced by
approximately a factor of two compared to a LO calculation. The resulting error is still rather large, corresponding
to approximately $+40$\% and $-30$\% across the region of Higgs masses, $150$~GeV~$< m_H < 180$~GeV.

For the LHC we have provided a brief study of the behaviour of our predictions for collisions at $\sqrt s = 10$~TeV.
We have also performed an analysis of this channel in the context of detecting a Higgs boson via weak boson
fusion, where the improved theoretical prediction presented in this paper is essential in the long-term for making a
measurement of the Higgs boson couplings to $W$ and $Z$ bosons.

\acknowledgments

We would like to thank Babis Anastasiou, Massimiliano Grazzini and Giulia Zanderighi for useful discussions.
CW acknowledges the award of an STFC studentship. Fermilab is operated by Fermi Research Alliance, LLC under
Contract No. DE-AC02-07CH11359 with the United States Department of Energy.

\end{document}